# Broad frequency and amplitude control of vibration in freestanding graphene via scanning tunneling microscopy with calculated dynamic pseudo-magnetic fields

P. Xu, J. K. Schoelz, S. D. Barber, M. L. Ackerman, and P. M. Thibado[a]

*Department of Physics, University of Arkansas, Fayetteville, Arkansas 72701, USA*

A technique to locally generate mechanical vibrations in freestanding graphene using scanning tunneling microscopy (STM) is presented. The frequency of the mechanical vibrations is tuned over nearly four decades and is centered around 10 Hz. The amplitude of the vibrations also changes over nearly three decades centered on 1 nm. The oscillating motion is generated in two ways: first, by scanning the STM tip on the surface and second, by scanning the bias voltage on the STM tip. The frequency and amplitude of the displaced freestanding graphene is quantitatively transformed to the frequency and strength of the locally generated pseudo-magnetic field for our specific geometry.

---

[a] Electronic mail: thibado@uark.edu

I.   INTRODUCTION

Graphene is a 2D system comprised of a single sheet of $sp^2$–bonded carbon atoms, and it has remarkable mechanical features in addition to its celebrated and rare electronic properties. Graphene is especially interesting considering that it was originally thought to be mechanically unstable. Since this misconception, it has been discovered that graphene exists via stabilizing ripples[1,2] and now holds the record for strongest material.[3]

The discovery of graphene was made even more exciting when mechanical distortions were found to alter its electromagnetic characteristics. For example, when uniaxial strain is applied to the honeycomb lattice, it affects the hopping probability for electrons in the $p_z$ orbitals. The resulting energy change can be represented mathematically by a position-dependent vector potential, as described by Castro Neto *et al.*, thereby suggesting the presence of a magnetic field.[4,5] Further analysis clarified that it is in fact a pseudo-magnetic field because opposite Dirac cones are related in such a way as to preserve global time-reversal symmetry. Guinea *et al.*[6] explained that this field should, nevertheless, still have a measurable impact. Verification came from Levy *et al.*,[7] who grew graphene on a Pt substrate. After cooling to ~7.5 K the graphene formed triangular nanobubbles due to its negative thermal expansion coefficient.[8] Scanning tunneling spectroscopy (STS) was then performed on the nanobubbles, and a series of peaks in the spectra was attributed to Landau levels originating from static magnetic fields with magnitudes as high as 300 T.

In parallel with this work, several techniques have been developed to investigate the basic mechanical properties of graphene. In one particular instance, Lee *et al.* deposited a graphene flake onto a Si substrate patterned with an array of circular wells.[3] Measurements made using an atomic force microscope permitted the Young's modulus of graphene to be calculated. In another study, Bunch *et al.* suspended graphene flakes over long rectangular trenches in a $SiO_2$ substrate, and used two unique methods to drive mechanical vibrations.[9] Their study involved applying an oscillating voltage to the graphene sheet, then modulating the intensity of a laser focused on the freestanding graphene to drive periodic expansion and contraction. The amplitude of the oscillations was measured (in the pm range) as a function of driving frequency (in the MHz range) using reflected light from a second laser, which allowed the resonant frequency and quality factor to be determined.

Overall, both the electronic nature and the mechanical movement of graphene could be brought together to generate controlled dynamic pseudo-magnetic fields. In this study, we demonstrate controlled mechanical vibrations in freestanding graphene using two scanning probe techniques. Nearly four decades of frequency control between mHz and Hz are demonstrated. Furthermore, nearly three decades of amplitude control between pm and nm are also achieved. Finally, quantitative predictions of induced dynamic pseudo-magnetic fields of up to 25 T are presented.

## II. EXPERIMENT

Experiments were performed using an Omicron ultrahigh-vacuum (base pressure is $10^{-10}$ mbar), low-temperature scanning tunneling microscopy (STM) operated at room temperature. Graphene layers grown using chemical vapor deposition[10] were transferred from Ni onto a 2000-mesh, ultrafine Cu grid with a square lattice of holes measuring 7.5 μm wide and bar supports 5 μm wide.[11] The grid was then mounted onto a flat tantalum sample plate using silver paint and transferred into the STM chamber, where it was electrically grounded. STM tips were manufactured in-house by electrochemically etching 0.25 mm diameter polycrystalline tungsten wire via a custom double lamella method with an automatic gravity cutoff.[12] After etching, the tips were gently rinsed with distilled water, briefly dipped in a concentrated hydrofluoric acid solution to remove surface oxides,[13] and transferred through a load-lock into the STM chamber. Filled-state STM images of freestanding graphene were acquired using a tip bias of +0.1 V and a constant tunneling current of 1 nA.

### III. RESULTS AND DISCUSSION

#### A. Frequency and amplitude control by scanning the tip

Four height line profiles, each 5 nm long, are displayed in Fig. 1(a). These are cross-sections taken from 12 nm × 12 nm STM images of freestanding graphene. The lines are plotted with offsets and in order of scan speed (indicated in the figure) used while acquiring corresponding STM image. The speed was roughly halved each time from about 72 nm/s at the top to 9 nm/s at the bottom. Notice that the wavelength, which is determined solely by the spatial periodicity of the honeycomb structure of graphene, is the same in each case, as expected. More surprisingly, as the scan speed decreases, the height of the oscillations increases. Starting at the top and working down, the average corrugation amplitudes (peak-to-peak) are approximately 0.13 nm, 0.45 nm, 0.90 nm, and 1.56 nm. All of these are significantly larger than the expected electronic corrugation of 0.05 nm for graphene.[14]

The time required for the STM tip to scan along a given height line profile is shown in Fig. 1(b). For each line profile shown in Fig. 1(a), the position axis value was divided by the appropriate scan speed in order to find the time in seconds. Since the top line profile was taken with the highest scan speed, this line profile is now shorter than the others. The remaining profiles each take about twice as long as the previous because the scan speed was divided by two every time. The frequency of the oscillations was determined by measuring the average peak-to-peak distance for each profile and is labeled in the figure. The frequencies, from top to bottom, are approximately 220 Hz, 108 Hz, 56 Hz, and 28 Hz. An STM image of freestanding graphene measuring 0.9 nm × 0.9 nm with honeycomb resolution is shown as an inset in Fig. 1(b). This was cropped from one of the larger images and is displayed with a compressed color scale to show the origin of the line profiles.

In addition to the established frequency control, the amplitude can also be tuned. The large vibration heights observed in this study have two distinct origins. The first is the spatial variation in the electronic density of states, which is normally observed with STM and known to be at most 0.05 nm.[14] The second mechanism, accounting for the balance of the movement, is the local elastic distortion caused by a varying electrostatic attraction between the tip and sample.[11] To help illustrate this interaction, a schematic of the STM tip moving over the freestanding graphene is provided in Fig. 1(c). If the STM tip is brought into tunneling above a carbon atom, as shown on the left side of Fig. 1(c), the graphene will be attracted to the tip and move toward it. In order to maintain a constant current, the tip must retract, but the graphene will follow the tip until an elastic restoring force equal to the electrostatic force develops. When the tip then scans away from the atom and moves over a hole, the electrostatic attraction will decrease,[11] and the graphene sheet will pull away from the tip, reducing the tunneling current and causing the tip to chase the graphene. Eventually, a new equilibrium configuration will be reached, with the graphene in a more relaxed state, as illustrated on the right side of Fig. 1(c). This repetitive, dynamic, and interactive process is in perfect registry with the electronic corrugations, greatly enhancing the vertical STM tip movement when imaging freestanding graphene. The height of the vibrations systematically increases as the scan speed decreases, because with lower speeds the system has ample time to reach equilibrium, thereby allowing the corrugation amplitude to grow to its maximum size. Thus, in summary, by scanning the STM tip over the surface and obtaining honeycomb resolution, the freestanding graphene can be forced to vibrate over a broad frequency range and with various amplitudes.

**B. Frequency and amplitude control by scanning the bias**

The second scanning probe technique used to drive the freestanding graphene is called electrostatic manipulation STM or EM-STM.[15] This measurement is carried out by moving the tip to a chosen location on the freestanding graphene and scanning the bias voltage on the STM tip while leaving the feedback circuit on. With the setpoint current fixed at 1 nA the height change of the STM tip gets recorded.[11] This technique is very similar to constant-current scanning tunneling spectroscopy (CC-STS), but whereas CC-STS is a spectroscopic measurement, EM-STM is primarily a mechanical measurement. Four EM-STM measurements, all under the same conditions and obtained in sequence, are shown in Fig. 1(d). Here the tip height is plotted as a function of bias voltage on the tip. The results are similar to each other, and as the voltage increased from 0.1 V to about 1.0 V, the height of the tip generally increased linearly by about 25 nm. The linear behavior continued from 1.0 V to 3.0 V, except with a smaller slope, with the tip height increasing only by another 5 nm. The measured tunneling current as a function of the tip bias for one of the EM-STM measurements is plotted as an inset in Fig. 1(d). The tunneling current remained constant at the setpoint value of 1.0 nA, confirming that the suspended graphene sample was indeed moving with the tip during the displacements.

The four EM-STM results were all acquired sequentially using the same bias voltage ramp rate of 2.04 V/s. Using this ramp rate the four profiles are plotted as a function of time to produce a saw-tooth type wave form as shown in Fig. 1(e). The period of the wave is 1.42 s, yielding a driving frequency of 700 mHz for the freestanding graphene. Note that other frequencies can be accessed by simply modifying the voltage ramp rate, and a different maximum amplitude may be set by changing the maximum voltage or even the setpoint of the tunneling current.[11]

A cross-sectional illustration showing the freestanding graphene below an STM tip for the EM-STM measurement is shown in Fig. 1(f). The tip is at a low bias voltage and there is little electrostatic attraction between the tip and the grounded sample for the schematic on the left. As the voltage is increased, however, the electrostatic attractive force will increase between the tip and sample, drawing it toward the tip, which must then retract in response to maintain the constant tunneling current. Equilibrium is reached when the elastic restoring force in the graphene balances the electrostatic force, as illustrated on the right side of Fig. 1(f).

After combining all the results using the two scanning probe techniques, the measured driving frequencies of the freestanding graphene are shown as a function of scan speed on a log-log plot in Fig. 2(a). Filled squares represent the STM tip scanning across the surface data, while the one open square represents the bias voltage being scanned. A linear regression was applied to the filled square data and has a slope equal to a reciprocal wavelength (i.e., $1/0.32$ nm$^{-1}$), allowing the open square data to be placed on the trend line. The total frequency range spans almost four decades, and it would be relatively easy to extend the range by another decade on either end. In a broader sense, the combination of EM-STM for slow oscillations and STM for rapid oscillations provides a method to locally and precisely generate dynamic curvature in freestanding graphene.

The measured corrugation amplitudes of the freestanding graphene also cover a wide range as a function of the scan speed as shown in Fig. 2(b). Again the filled squares represent the STM data, while the single open square represents the EM-STM data. The height of the oscillations systematically decreases as the scan speed increases as discussed earlier. A simple polynomial running through all the data points shows that the amplitude values in our measurements span nearly 3 decades. This range could also be extended by altering the experimental settings.

**C. Vibration-derived dynamic pseudo-magnetic fields in freestanding graphene**

The ability to drive the freestanding graphene in a controllable manner may be utilized to further study the relationship between the mechanical, electronic, and pseudo-magnetic properties of graphene. For example, in a previous study, Xu et al., used the analytical expressions for the deformation of a membrane from Ref. 16, along with expressions for the pseudo-magnetic field in Ref. 6 and the deformation potential in Ref. 17, to estimate the pseudo-magnetic field created by the induced strain. For this same freestanding graphene geometry and for small displacements, they found the pseudo-magnetic field varied linearly at a rate of about 0.7 T/nm.[11] Using this result, we calculated that the pseudo-magnetic field oscillates in space with amplitude varying from about 0.5 T to 2 T while the graphene is vibrating in the patterns shown in Fig. 1(a). Similarly, the pseudo-magnetic field also oscillates in time with the same field strength (0.5 T to 2 T) for the patterns shown in Fig. 1(b). In order to generate larger fields over longer time periods the EM-STM technique can be used. The pseudo-magnetic field increases to a size of about 25 T, as displayed on the right side of Fig. 1(d). Furthermore, the rate of increasing pseudo-magnetic field is demonstrated to be as high as 100 T/s, as shown on the right side of Fig. 1(e). All together, the magnitude of the pseudo-magnetic field can be tuned through a broad range using both STM and EM-STM as shown on the right side of Fig. 2(b). Here the field strength can be tuned from about 1 T to 100 T. Furthermore, as the strength of the field varies, so does its frequency. By combining the results in Fig. 2(a) with those in Fig. 2(b), we demonstrate that the strength of the pseudo-magnetic field decreases with increasing field frequency in Fig. 2(c). Specifically, as the frequency of the field increases from 1 Hz to a few hundred Hz the field strength decreases from 100 T to 0.1 T.

There are several fascinating points related to these results that need to be discussed. One is the relatively small displacements required to generate such large pseudo-magnetic fields. This leads to fast switching times and low energy costs. Also, the large range of frequency and amplitude control provides access to numerous ways to test and modify experimental outcomes. It is also possible to consider using these dynamic field effects near defects and edges already present in freestanding graphene, in order to test if time-reversal symmetry may be broken, which is an important basic physics question for massless, Dirac fermions.[18]

## IV. SUMMARY AND CONCLUSIONS

In summary, we have demonstrated frequency-controlled dynamic manipulation of freestanding graphene using STM and EM-STM. Nearly four decades of frequency control centered at 10 Hz has been demonstrated. In addition, we showed it is possible to tune the amplitude of the oscillations across nearly three decades centered around 1 nm. The scanning speed of the STM tip on the surface and the scanning speed of the bias voltage on the tip are shown to control the frequency of the vibrating membrane. Finally, we quantified the magnitude of the pseudo-magnetic field generated for our geometry and showed how this field is locally generated and tuned over the same frequency range.

**ACKNOWLEDGEMENTS**

Special thanks are given to S. Barraza-Lopez for his insightful comments. The authors gratefully acknowledge the financial support of the Office of Naval Research under grant number N00014-10-1-0181 and the National Science Foundation under grant number DMR-0855358.

**Figure Captions**

FIG. 1. (a) Four height cross-section line profiles taken from STM images of freestanding graphene and acquired with indicated scan speeds. (b) The same line profiles from (a) divided by their respective scan speed converts the *x*-axis to time. Inset: Honeycomb resolution filled-state STM image of freestanding graphene. (c) Schematic of the STM tip scanning over the graphene surface when above a carbon atom (left) or a hole (right). (d) Four sets of EM-STM data acquired on suspended graphene with a constant setpoint current of 1.0 nA. Inset: Measured tunneling current as a function of voltage acquired during EM-STM. (e) The EM-STM data is converted to show the height of the tip as a function of time. (f) Schematic of the EM-STM measurement on graphene at a low-voltage (left) and a high voltage (right).

FIG. 2. (a) Measured frequency of the oscillating freestanding graphene plotted as a function of the scan speed and shown with a linear fit. One EM-STM data point is also included along the fit line. (b) Vibration amplitudes of freestanding graphene are shown as a function of the scan speed. One EM-STM data point is also included. (c) Calculated magnitude (or strength) of the pseudo-magnetic field shown as a function of the oscillating frequency of the pseudo-magnetic field.

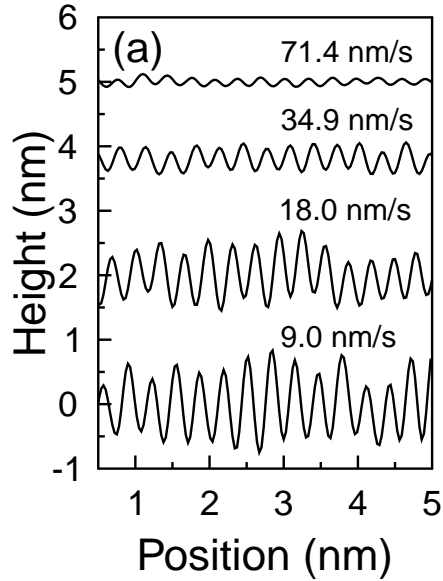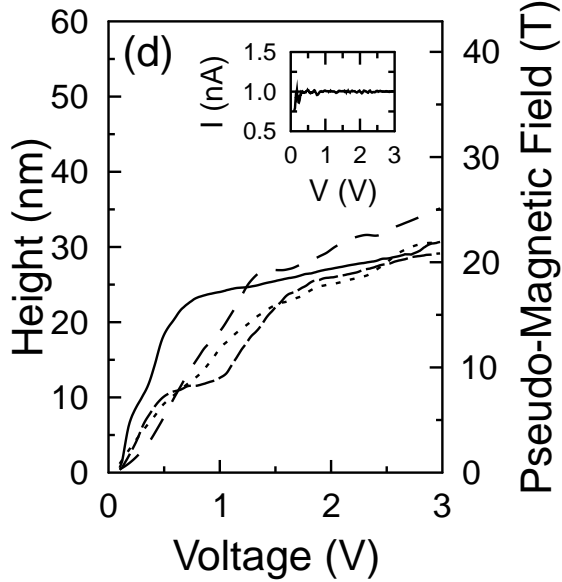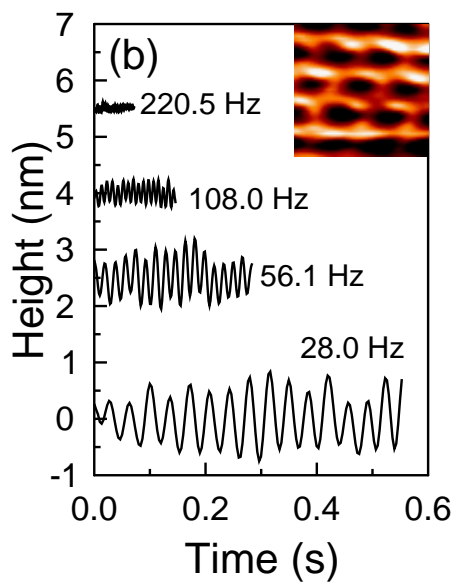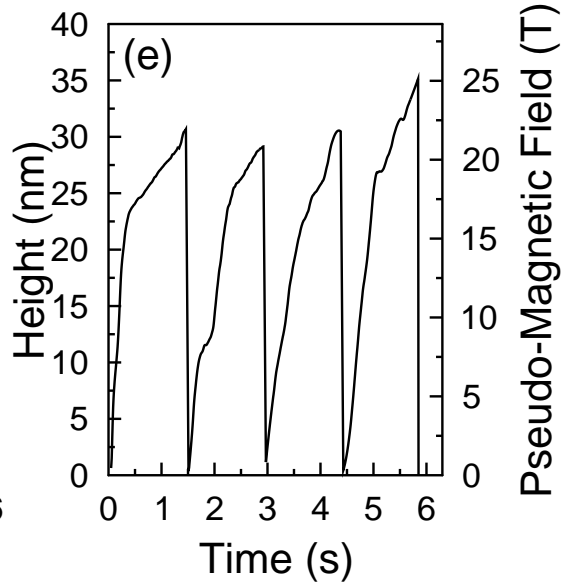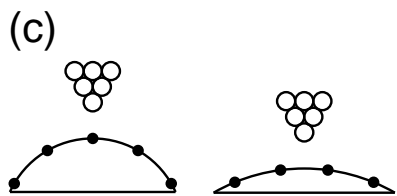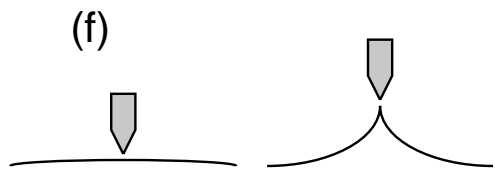

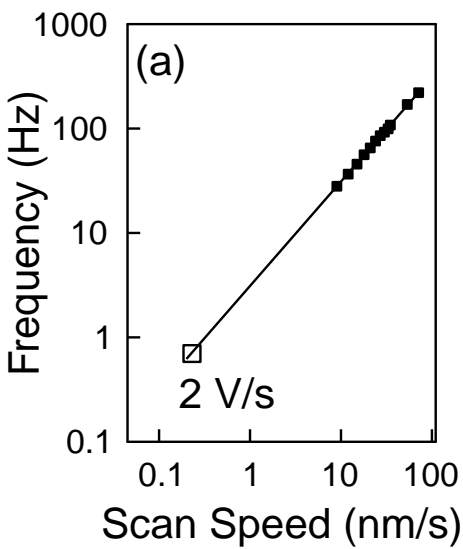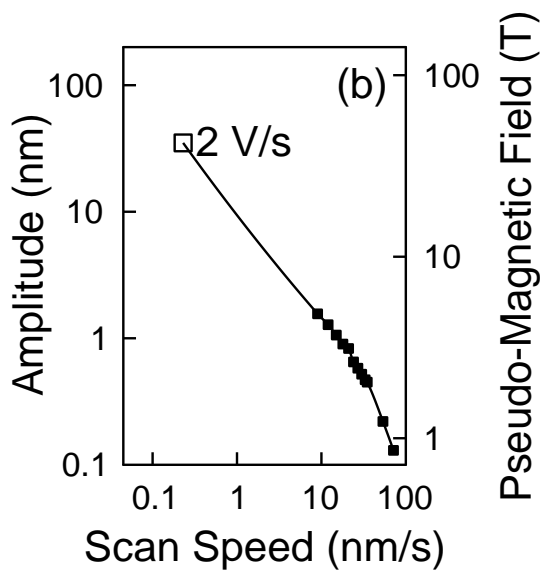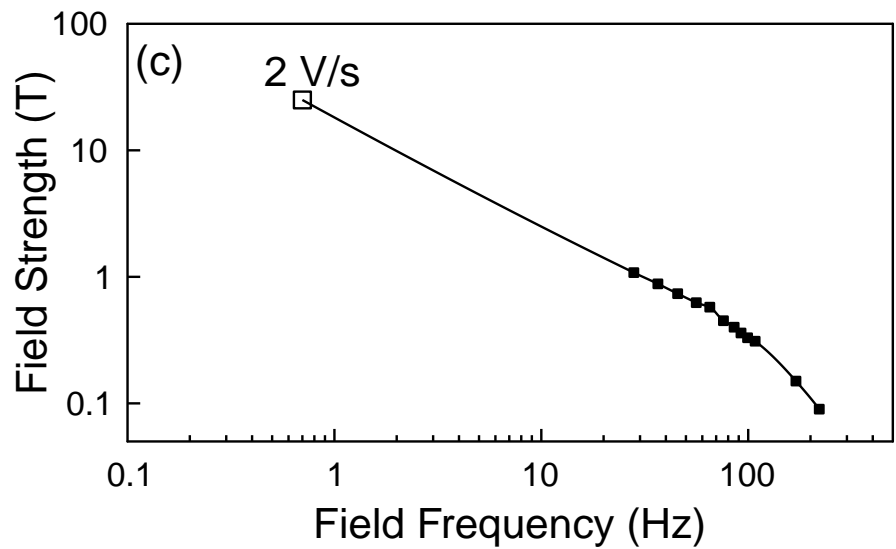